\documentstyle[12pt]{article}
\def\by#1#2{{\displaystyle {#1}\over \displaystyle {#2}}}

\def\gev2{\hbox{GeV}^2}
\begin{document}

\begin{flushright}
DO-TH-96/17 \\
September 1996
\end{flushright}

\begin{center}
{\Large \bf On the $Q^2$ dependence of Nuclear Structure Functions} \\
[0.8cm]
D.\ Indumathi, \\ 
[0.3cm]
{\it Institut f\"ur Physik, Universit\"at Dortmund, D 44221, Dortmund,
Germany} \\ [1cm]
\end{center}

\begin{abstract}


The recent high statistics NMC data on the Tin to Carbon structure
function ratio seems to indicate, for the first time, a significant $Q^2$
dependence, especially at small values of Bjorken $x$, $x < 0.05$, and
$Q^2 > 1\ \gev2$. Pure leading twist, perturbative QCD--based
predictions, which are consistent with the free nucleon data, yield a
fairly flat ratio with little or no $Q^2$ dependence. In view of this
seeming contradiction, we re-examine the applicability of such a
perturbative model to nuclear structure functions in such a kinematical
regime. We find that the model is consistent with all data, within
experimental errors, without any need for introducing additional higher
twist contributions. The model correctly reproduces the $Q^2$ dependence
of the Carbon structure function as well. We also critically examine
the $Q^2$ dependence of the corresponding spin dependent structure
functions. 

\end{abstract}

\vspace{0.5cm}

\section{Introduction}

The structure function, $F_2^A(x, Q^2)$, for various nuclei of mass
number, $A$, such as
${}^4$He, ${}^6$Li, ${}^{12}$C, ${}^{40}$Ca, and ${}^{208}$Pb, has
been precisely measured in several lepton--nucleon deep inelastic
scattering (DIS) experiments \cite{NMC,E665}, over a large
range of the Bjorken variable, $x$, and momentum transfer, $Q^2$. In
all cases, the ratio of the structure function for the bound nucleon
in a nucleus, $A$, to that for the free nucleon, $D$ (taken to be an
``average'' nucleon in Deuteron), $R^A = F_2^A/F_2^D$, was found to be
less than unity at small and large values of $x$, with antishadowing
($R > 1$) at intermediate $x$ values around $x = 0.1$. No significant
$Q^2$ dependence of these ratios has been observed so far \cite{NMC,E665},
within the error bars of these measurements.
That is, the nuclear (bound nucleon) and free nucleon structure
functions seem to exhibit the same $Q^2$ dependence.

A recent measurement \cite{NMCQ2} by the NMC presents, for the first
time, a high-statistics study of the $Q^2$ behaviour of the Tin to
Carbon structure function ratio, $F_2^{\rm Sn}/F_2^{\rm C}$;
consequently, the statistical errors are much smaller than in the
earlier measurements. The data appear to show significant $Q^2$
dependences, unlike what has been observed before, for other nuclei.
Several theoretical models of nuclear structure functions are available
\cite{MT,Q2dep,Kumano}, that can account for such a $Q^2$ dependence,
especially at small values of $x$ and $Q^2$. Many of these use a
hadronic approach to the problem, for example, vector meson dominance
(VMD), which is valid down to nearly the photoproduction limit, and
Regge phenomenology; this results in effectively a higher twist (or a
$1/Q^2$ type) contribution, that can reproduce the observed large,
positive slope of the structure function ratio at small values of
$x$, $x < 0.05$ \cite{MT}. As has been pointed out \cite{MT}, a
perturbative approach to the problem, addressed, similar to the free
nucleon case, via the partonic content of the bound nucleon in the
leading twist (LT) approximation, would result in small or essentially
no $Q^2$ dependence of structure function ratios; such a partonic
approach should also be more applicable at larger $Q^2$. A perturbative
approach, in principle, also includes a higher twist (HT) contribution,
that dies away as $1/Q^2$. Such a contribution would therefore not be
significant for $Q^2$ much larger than, say, $10\ \gev2$. There is
no compulsive evidence for such HT contributions even down to
$1 \ \gev2$ in the free nucleon case \cite{GRV}. However, it is not
known, at present, whether the $Q^2$ dependence of free and bound
nucleon structure functions are the same. The current Sn/C
data at small $x$, which show a significant $Q^2$ dependence, are
precisely in the region of $1 < Q^2 < 10 \ \gev2$ where it is not
clearly established that HT effects are subleading. It is pertinent,
therefore, to ask what is the smallest $Q^2$ down to which a pure LT
contribution to bound nucleon structure functions is still consistent
with the new Sn/C measurement. The answer, surprisingly enough, is that
a pure DGLAP approach is still consistent with {\it all} available data,
within the experimental errors. We now present the details of the
calculation, and substantiate our claim.

\section{Brief description of the model}

Specifically, we choose, for the purposes of effecting this
comparison, a model \cite{IZ} that uses standard leading order DGLAP
\cite{DGLAP} (or leading twist) evolution equations to obtain the
nuclear structure functions at various $Q^2$ values, starting from a
low input scale, $Q^2 = \mu^2 = 0.23\ \gev2$. The Gl\"uck-Reya-Vogt
(GRV) \cite{GRV} parametrisation is used for the free nucleon input,
which is well-known to provide a good fit to the available proton data;
the corresponding bound nucleon input densities are obtained by
calculating distortions of the free ones due to the effects of nuclear
binding and swelling. Binding, in particular, plays a special r\^ole at
low-$x$, where shadowing is observed. Details of the model can be found
in Ref.~\cite{IZ}. In summary, the increased radius of the bound
nucleon, due to nucleon swelling \cite{Jaffe}, results in a
redistribution of the partons inside the nucleon \cite{Zhu}. The
increase in the radius, $\delta_A$, of a bound nucleon, relative to a
free one, is parametrised as,
$$
\delta_A = \delta_{\rm vol} (1 - 1.1 A^{-1/3})~;
\eqno(1)
$$
the free parameter, $\delta_{\rm vol}$, was chosen to be 15\%. Eq. (1),
therefore, fixes the swelling for all $A$. Conservation of parton number
and energy--momentum, along with use of Heisenberg's uncertainty
principle, then specifies the input bound nucleon parton densities,
starting from the free nucleon ones, using only one free parameter
that parametrises the extent of swelling, and with fixed $A$ dependence,
as indicated in eq.~(1). Nuclear binding is introduced by assigning the
consequent energy loss, as computed using the Weizs\"acker mass formula,
to the sea quark sector; the sea quark density is thus depleted in
comparison to the free nucleon case, by an amount dependent on its mass
number, $A$. Such a depletion also occurs at the time of scattering,
for partons with small $x$, $x < 0.1$ \cite{IZ}. The model predicts
structure functions with very few free parameters, and with well-defined
$A$ dependences. Results have been obtained for various structure
function ratios at average values of $Q^2$, corresponding to the
measured values; good agreement with data has been found in the region
of validity of the model. (The model is not valid at large $x$,
$x \,{\stackrel{\displaystyle {}_>}{\displaystyle {}_\sim}} \,0.5$,
since Fermi motion effects have not been included). 

\subsection{Comparison with data}

\begin{figure}[ht]

\vskip20truecm

\includegraphics{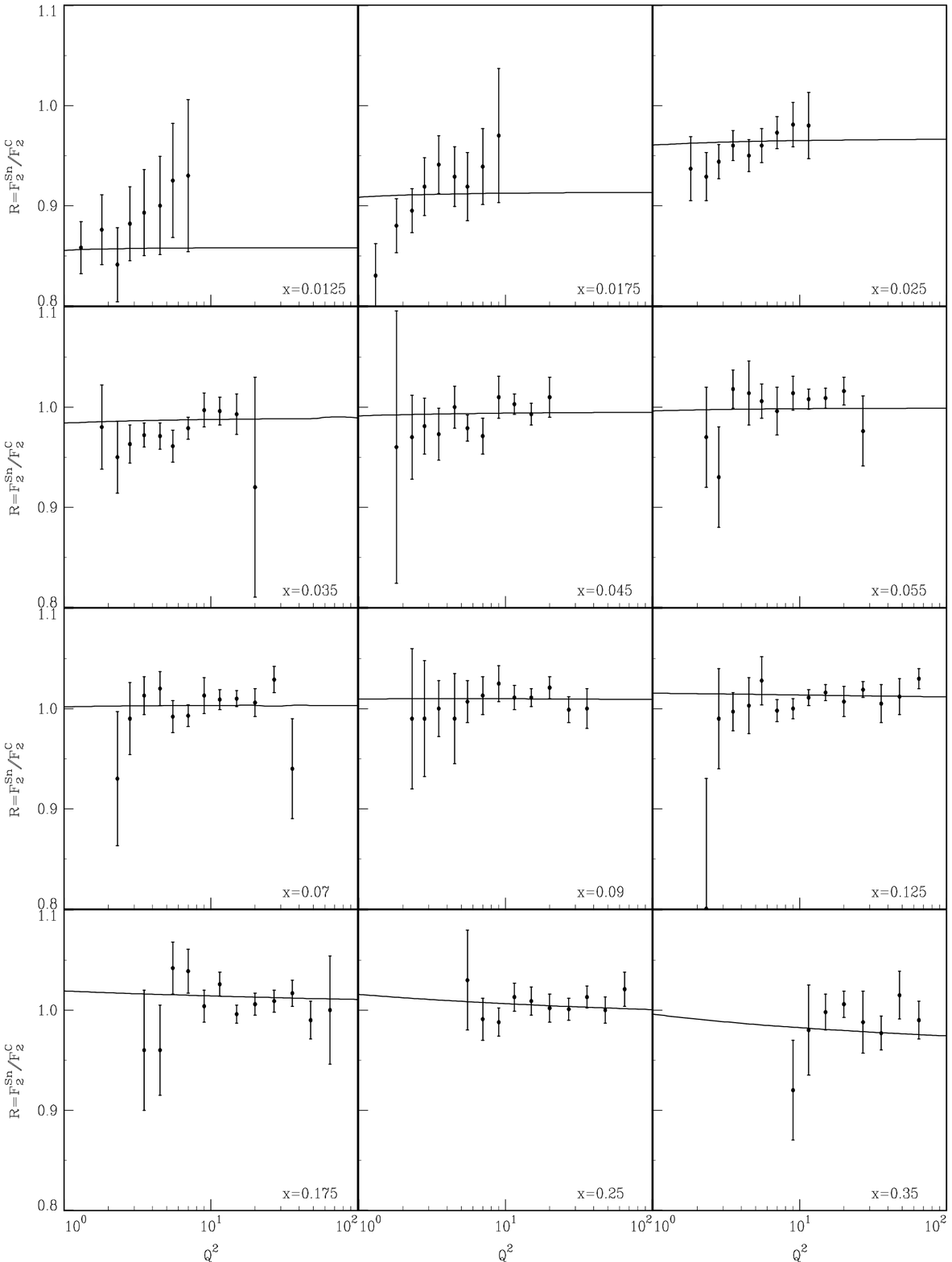}
\caption[dummy]{\small The model prediction for the $Q^2$ dependence of
the structure function ratio for Sn/C, in comparison with data from the
NMC \cite{NMCQ2}, with statistical and systematic errors added in
quadrature. Average (central bin) values of $x$ are shown.}
\label{fig1}
\end{figure}

In view of the new Sn/C data, we now examine in detail the $Q^2$
dependence of
the ${}^{118}$Sn and ${}^{12}$C nuclear structure functions and their
ratios within this dynamical model. We emphasise that, once the model
has been fitted to data from nuclei such as Helium or Carbon, as has
already been done, the resulting Tin structure function is essentially
fixed, with no more free parameters. Furthermore, the magnitude and the
$Q^2$ dependence in the base
case, viz., for the structure function of proton or deuteron (which is
considered to be the prototype ``average nucleon'') is well reproduced
by this model. (It is in fact tailored to do so). Since the bound
nucleon densities are computed in a manner similar to that of the free
nucleon ones, that is, by the DGLAP evolution equations, the $Q^2$
dependence of the bound densities is similar to that of the free ones.
The only difference lies in the input (starting) densities, which are
computed according to standard nuclear inputs. Hence our model exhibits
only a weak $Q^2$ dependence for all structure function ratios
\cite{IZ2}. This can be seen in Fig.~\ref{fig1}, where the model
prediction for the structure function ratio, $R = F_2^{\rm Sn}/F_2^{\rm
C}$, is plotted for various $x$ values in the range between about 0.01
and 0.4 (the central value of the corresponding experimental $x$ bin
was used) as a function of $Q^2$, for $1 < Q^2 < 100$ $\gev2$. Note that
all the data lie in the accepted perturbative regime of $Q^2 > 1
\gev2$. 

Although the central values of the data at low-$x$ are strongly increasing
with $Q^2$, it is seen that the model is in good agreement with the
available data, within the errors of the experiment. The $\chi^2$ per
degree of freedom of the fits, as shown in Table~\ref{table1}, is seen
to be around unity or less. The only exception is the last $x$-bin,
$x = 0.55$; however, here our model is not expected to hold, since we
have not included Fermi motion effects. (This also seems to indicate
that these effects are severely $A$-dependent and so do not cancel in
the ratio of structure functions). The average $\chi^2/$d.o.f. $= 138/142$
for the complete data set, indicating that the model is consistent with
the data. This further improves to 109/136 if we drop the last data
point, $x = 0.55$. 

\begin{table}[ht]
\begin{center}
\begin{tabular}{|l|c|c||l|c|c||l|c|c|}
\hline
\quad $x$ & $\chi^2$/d.o.f. & d.o.f. & 
\quad $x$ & $\chi^2$/d.o.f. & d.o.f. & 
\quad $x$ & $\chi^2$/d.o.f. & d.o.f. \\
\hline
0.0125 &  0.59 & 8  & 0.055  &  0.78 & 11 & 0.25   &  0.53 & 11 \\ 
0.0175 &  1.18 & 9  & 0.07   &  0.93 & 12 & 0.35   &  1.10 & 9  \\
0.025  &  0.65 & 9  & 0.09   &  0.27 & 12 & 0.45   &  1.16 & 7  \\ 
0.035  &  0.96 & 11 & 0.125  &  0.90 & 14 & 0.55   &  4.77 & 6  \\  
0.045  &  0.55 & 11 & 0.175  &  0.99 & 12 &        &       &    \\
\hline
\end{tabular}
\caption[dummy]{\small The $\chi^2$ fits of the model with the NMC data
on the $Q^2$ dependence of the Tin to Carbon structure function ratio
\cite{NMCQ2}, for various $x$ bins, the central values of which are
shown. }
\label{table1}
\end{center}
\end{table}

Hence, we see that a purely DGLAP-type perturbative evolution of an
input set of densities predicts a very small $Q^2$ dependence of
the Sn/C structure function ratio \cite{MT,Q2dep}. Moreover, the
prediction is consistent with the available NMC data.

The $Q^2$ dependence of the
structure function ratios, $R^{A} = F_2^A/F_2^D$, for $A$ = Li, C, Ca,
and Pb, also show very little $Q^2$ dependence in the shadowing regime,
$x \le 0.1$. The resulting fit to a linear $Q^2$ dependence,
of the form,
$$
\by{F_2^A}{F_2^D} = b\,\ln Q^2 + c~,
\eqno(2)
$$
yields a slope, $b$, that, though positive at small-$x$, and negative at
large $x$ $(x > 0.1)$, is almost vanishing throughout, and compatible
with the corresponding measurements available from NMC and E665
\cite{NMC,E665} in the region of overlap (we do not include the points
which have $x > 0.5$ and $Q^2 \le 1\ \gev2$ since we are using a
perturbative evolution equation without Fermi motion effects). We
therefore believe that the model predictions are compatible with the
available data. 

\begin{figure}[ht]

\vskip20truecm

\includegraphics{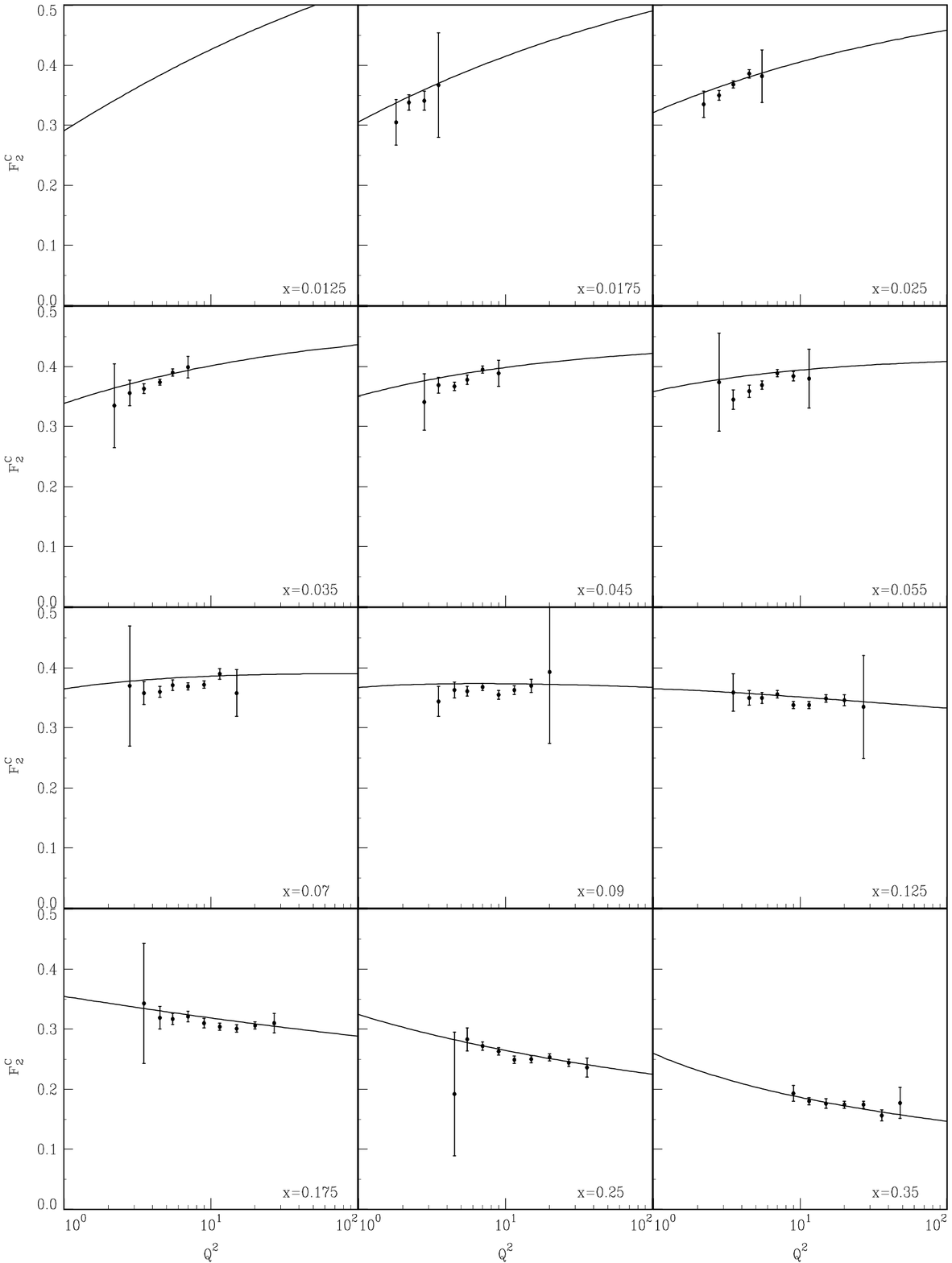}
\caption[dummy]{\small The model prediction for the $Q^2$ dependence
of the bound nucleon structure function in Carbon for various values
of $x$. Data shown correspond to preliminary results from the NMC
\cite{Mucklich}, with only statistical errors shown.}
\label{fig2}
\end{figure}

This implies that the $Q^2$ dependence of the bound and free structure
functions are essentially the same. Note that the corresponding
{\it free} nucleon data in the same $x$ region
can be fitted with a DGLAP type $Q^2$ behaviour and do not seem to need
inclusion of HT effects in the region $1 < Q^2 < 10 \ \gev2$. Hence the
$Q^2$ dependence of free and bound nucleon structure functions seems to
show the same origin, viz., parton splitting. We test this by comparing
the $Q^2$ dependence of the Carbon structure function data alone, not
just structure function ratios, against the preliminary high statistics
data available from the NMC \cite{Mucklich}. 
This is shown in Fig.~\ref{fig2}, for the same ranges of $x$, and for
the same $Q^2$ bins as the Sn/C data. Note that the error bars shown
here correspond to statistical errors alone. We see that there is
reasonable agreement with the data for all $x$ ranges. However, the
$\chi^2$/d.o.f. here is worse, being only 122/87; this is because of
worse fits for $0.04 \le x \le 0.1$ as can be seen from
Table~\ref{table2}. This can be greatly improved by an overall
normalisation change and does not require a change in the
slope\footnote{This can be accomplished by a small change in the extent
of swelling, which remains a free parameter in this model. This would
cause a corresponding change in the structure functions of all other
nuclei; however, this should not worsen the agreement of the model with
these other measured structure functions.}. For instance, a decrease in
the overall normalisation by just 3\% decreases the $\chi^2/$d.o.f. to
74/87 (or to 68/83, if we drop the last data point), which is very
acceptable. However, we do not do this here, and merely keep in mind
that there would also be an improved agreement on using a
next-to-leading (NLO) order fit rather than just a leading order one. 

\begin{table}[ht]
\begin{center}
\begin{tabular}{|l|c|c||l|c|c||l|c|c|}
\hline
\quad $x$ & $\chi^2$/d.o.f. & d.o.f. & 
\quad $x$ & $\chi^2$/d.o.f. & d.o.f. & 
\quad $x$ & $\chi^2$/d.o.f. & d.o.f. \\
\hline
0.0125 &  --   & -- & 0.055  &  2.95 & 7  & 0.25   &  0.79 & 9  \\
0.0175 &  0.64 & 4  & 0.07   &  2.70 & 8  & 0.35   &  0.43 & 7  \\
0.025  &  0.95 & 5  & 0.09   &  1.91 & 8  & 0.45   &  0.09 & 5  \\
0.035  &  1.26 & 6  & 0.125  &  1.19 & 9  & 0.55   &  1.35 & 4  \\
0.045  &  1.78 & 6  & 0.175  &  1.35 & 9  &        &       &    \\
\hline
\end{tabular}
\caption[dummy]{\small The $\chi^2$ fits of the model with the NMC data
on the $Q^2$ dependence of the Carbon structure function, $F_2^C(x,
Q^2)$ \cite{Mucklich}, for various available $x$ bins, the central
values of which are shown. }
\label{table2}
\end{center}
\end{table}

\subsection{The polarised case}

\begin{figure}[ht]

\vskip8truecm

\includegraphics{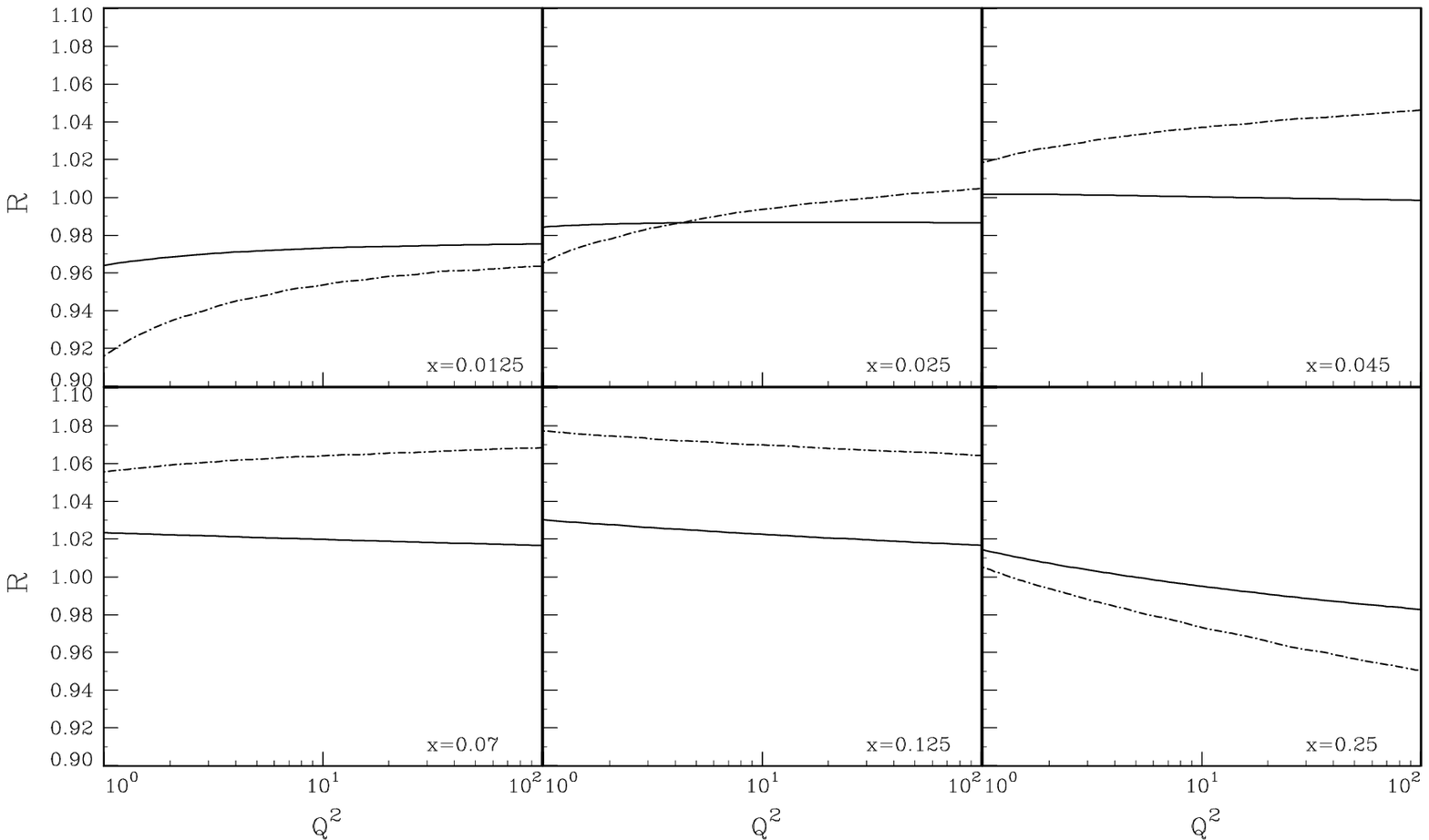}
\caption[dummy]{\small The model prediction for the $Q^2$ dependence of
the spin independent ($R$) and spin dependent ($\widetilde{R})$ structure
function ratios, for He/n, are shown as solid and dashed lines
respectively for various $x$ values.}
\label{fig3}
\end{figure}

Finally, we consider the $Q^2$ dependence of the corresponding
polarised structure function, $g_1^A (x, Q^2)$. In particular, the
structure function of interest \cite{pol} is that of ${}^3$He, since
it was the target used in determining the neutron spin dependent
structure function \cite{E142}. We use an extension of the same
dynamical model to the polarised case \cite{IZ,I} to study the
$Q^2$ dependence, where the free nucleon input now corresponds to the
fits from the Gl\"uck-Reya-Vogelsang (GRVs) \cite{GRVpol}
parametrisation. We see from Fig.~\ref{fig3} that the polarised ratio, 
$$
\widetilde{R} = \by{g_1^{\rm He/n}}{g_1^n}~,
\eqno(3)
$$
is more sensitive to $Q^2$ than the corresponding unpolarised ratio,
$$
R^n = \by{F_2^{\rm He/n}}{F_2^n}~,
\eqno(4)
$$
where the superscript, $n$, on the unpolarised ratio indicates that the
``average neutron'' structure function in He was computed, in order to
effect a meaningful comparison with $\widetilde{R}$. (See Table
\ref{table3} for a comparison with data.) At small $x$, where
the sea densities dominate, the $Q^2$ behaviour of $\widetilde{R}$ and
$R$ are different, owing to the different $Q^2$ dependence of the
corresponding unpolarised and polarised splitting functions, $P_{ij}$
and $\Delta P_{ij}$. At larger $x$, from
$x \,{\stackrel{\displaystyle {}_>}{\displaystyle {}_\sim}} \,0.1$,
since $P_{qq} = \Delta P_{qq}$,
i.e., the nonsinglet unpolarised and polarised splitting functions are
the same, $\widetilde{R}$ and $R$ show the same $Q^2$ behaviour. 

\begin{table}[ht]
\begin{center}
\begin{tabular}{|l|c|c|c|c|c|}
\hline
\quad $\langle x \rangle$  & $\langle Q^2 \rangle$ & 
$A_1^{\rm He}$(data) & $A_1^{\rm He}$(model) &
$A_1^{\rm He, model}(Q^2)$ \\
 & & & & ~~$A_1(1)$ : $A_1(10)$  \\ \hline
0.0055 &  --  & \qquad \quad --     & \quad -- & $-0.022$ : $-0.020$ \\
0.0085 &  --  & \qquad \quad --     & \quad -- & $-0.030$ : $-0.027$ \\
0.0125 &  --  & \qquad \quad --     & \quad -- & $-0.039$ : $-0.034$ \\
0.025  & 0.96 & $~~\,0.066 \pm 0.111$ & $-0.063$ & $-0.063$ : $-0.054$ \\
0.035  & 1.10 & $-0.058  \pm 0.060$ & $-0.078$ & $-0.078$ : $-0.066$ \\
0.05   & 1.30 & $-0.095  \pm 0.045$ & $-0.096$ & $-0.098$ : $-0.080$ \\
0.078  & 1.60 & $-0.062  \pm 0.044$ & $-0.116$ & $-0.123$ : $-0.098$ \\
0.124  & 2.30 & $-0.137  \pm 0.048$ & $-0.127$ & $-0.142$ : $-0.110$ \\ 
0.175  & 2.70 & $-0.087  \pm 0.055$ & $-0.122$ & $-0.141$ : $-0.108$ \\ 
0.248  & 3.10 & $-0.020  \pm 0.072$ & $-0.098$ & $-0.117$ : $-0.086$ \\
0.344  & 3.40 & $~~0.029 \pm 0.114$ & $-0.046$ & $-0.064$ : $-0.037$ \\
0.466  & 5.20 & $~~0.030 \pm 0.241$ & $~~0.021$ & $~~0.002$ : $~~0.028$ \\
\hline
\end{tabular}
\caption[dummy]{\small The model prediction for the polarisation
asymmetry in Helium compared with data \cite{E142} at the same central
values of $(x, Q^2)$ as the data; errors shown are statistical and
systematic, combined. The $Q^2$ dependence of the computed
asymmetry is indicated by showing the asymmetry at $Q^2 = 1, 10 \ \gev2$
in the last column. }
\label{table3}
\end{center}
\end{table}

\begin{figure}[ht]
\vskip8truecm

\includegraphics{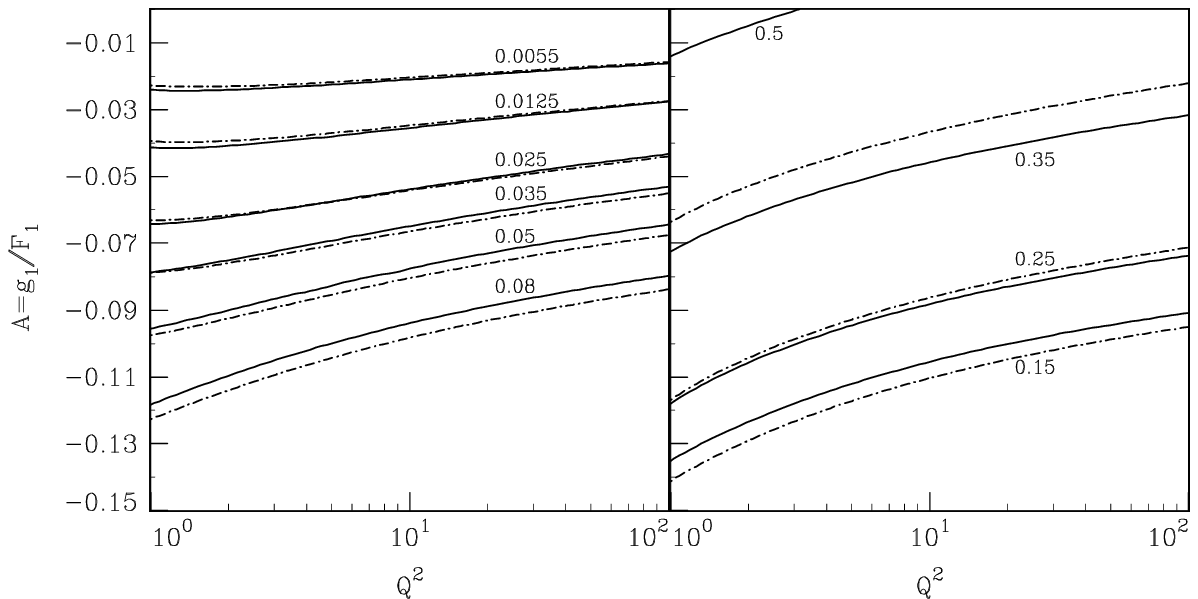}
\caption[dummy]{\small The model prediction for the $Q^2$ dependence of
the polarisation asymmetry, $A_1^{He}$ (dashed lines) and the free
neutron asymmetry, $A_1^n$ (solid lines), for various values of $x$ from
0.0055 to 0.5. A comparison with available data
\cite{E142} can be found in Table~\ref{table3}. }
\label{fig4}
\end{figure}

Such a $Q^2$ dependence \cite{GRVpol} implies that the corresponding
spin asymmetry,
$$
A_1^{A/n} \equiv \by{g_1^{A/n}}{F_1^{A/n}}~,
\eqno(5)
$$
which is what is usually measured in DIS with polarised targets, will
show more sensitivity to $Q^2$. This can be seen in Fig.~\ref{fig4}
where both the free ($A_1^n$) and bound ($A_1^{{\rm He}/n}$) nucleon
asymmetries are shown (using $F_2 = 2x F_1$) for various $x$ values
accessible by current experiments. The {\it extent} of the
$Q^2$ dependence, as has been exhibited by a comparison with the
available data in Table~\ref{table3}, gives hope that, when more data is
available, this large $Q^2$ dependence can actually be observed. For
instance, the asymmetry changes by about 15--20\% for $x$ from
0.025--0.2, in going from $Q^2 = 1$ to $Q^2 = 10 \ \gev2$. As has
already been pointed out \cite{I}, the nuclear dependence effectively
cancels in the ratio, eq.~(5), so that the free and bound nucleon
asymmetries are
nearly equal; hence the E142 measurement (which measures the bound
nucleon neutron asymmetry in Helium) still provides a good measure of the
free neutron asymmetry, and hence no information at all about nuclear
effects. It seems as if information on spin dependent {\it nuclear}
structure functions can only be obtained by measuring ratios similar to
that shown in Eq.~(3), or plotted in Fig.~\ref{fig3}:
$$
\widetilde{R}^{AB}(x, Q^2) = \by{g_1^A(x, Q^2)}{g_1^B(x, Q^2)}~,
\eqno(6)
$$
rather than from measurement of the
conventional polarisation asymmetry, $A_1^{A/n}$.

\section{Discussion and Conclusion}

We conclude that a perturbative approach to nuclear structure functions
with the use of the leading twist DGLAP evolution equations predicts,
not just the ratios, but also the magnitudes of various nuclear
structure functions, and their $Q^2$ behaviour as well, that are
consistent with all data so far available. Such an approach typically
yields a very flat $Q^2$ behaviour for structure function ratios at
small $x$. A similar $Q^2$ behaviour of bound and free nucleon spin
independent structure functions is therefore indicated. Another
perturbative approach, using parton recombination and $Q^2$ rescaling
arguments, has been able to obtain a non-trivial $Q^2$ dependence
\cite{Kumano}, similar in nature to that obtained by non-partonic
approaches \cite{MT,Q2dep}. The $Q^2$ dependence in this model, though,
is very sensitive to the input parameters \cite{Kumano}. However, a
persistence of the
significant $Q^2$ dependence in the Sn/C ratio at small $x$ (for
$Q^2 \ge 1 \ \gev2$), accompanied by a shrinking of the corresponding
error bars by the acquisition of more data, could well signify the
presence of higher twist effects, which have not so far been observed
for the free nucleon case at such $Q^2$ values. This would
significantly affect the extraction of spin dependent structure
functions from polarised deep inelastic scattering experiments off
polarised nuclear targets as well. Hence, a detailed experimental
study of the $Q^2$ dependence of the spin polarisation asymmetry is
also urged. In short, it would be interesting to obtain more data for
various nuclei in the small-$x$ region, both in polarised and
unpolarised experiments, in order to verify if a universal $Q^2$
behaviour of free and bound nucleon structure functions exists. This is
bound to throw more light on the nature and origin of the modifications
of parton densities in nuclei. 

\vspace{0.3cm}

\noindent {\bf Acknowledgements}: I am thankful to Micha\l\ Szleper for
kindly making the NMC data available to me, as well as to Michele
Arneodo for help with obtaining the data. I am grateful to W.~Zhu for
many discussions, and E.~Reya for constant encouragement, and critical
comments on the manuscript.


\begin{thebibliography}{99}

\bibitem{NMC}
P. Amaudruz {\it et al.}, The NMC, Nucl. Phys. B441 (1995) 3;
M. Arneodo {\it et al.}, The NMC, Nucl. Phys. B441 (1995) 12.

\bibitem{E665}
M.R. Adams, {\it et al.}, The E665 Collaboration, Z. Phys. C67 (1995)
403.

\bibitem{NMCQ2} Micha\l\ Szleper, private communication. 

\bibitem{MT} W. Melnitchouk and A.W. Thomas, Phys. Lett.
B346 (1995) 165; Phys. Rev. C52 (1996) 3373;
G. Piller, W. Ratzka, and W. Weise, Z. Phys. A352 (1995) 427.

\bibitem{Q2dep}
L.L. Frankfurt, M.I. Strikman, S. Liuti, Phys. Rev. Lett.
65 (1990) 1725; 
B. Kopeliovich, B. Povh, preprint, MPIH-V12-1995, to appear in Phys.
Lett.;
see also, \cite{Mucklich} below.

\bibitem{Kumano}
M. Miyama, S. Kumano, Phys. Rev. C50 (1994) 1247; Phys. Rev. C48
(1993) 2016; 
S. Kumano, M. Miyama, Phys. Lett. B378 (1996) 267; 
R. Kobayashi, S. Kumano, and M. Miyama, Phys. Lett. B354 (1995) 465.

\bibitem{GRV} M. Gl\"uck, E. Reya, and A. Vogt, Z. Phys. C48 (1990) 471;
Z. Phys. C67 (1995) 433.

\bibitem{IZ} 
D.~Indumathi, W. Zhu, Dortmund University preprint, DO-TH-95/03, 1995,
to appear in Z. f\"ur Physik {\bf C}.

\bibitem{DGLAP}
V.N. Gribov and L.N. Lipatov, Sov. J. Nucl. Phys. 15 (1972) 438;
{\it ibid}, 675;
Yu.L. Dokshitzer, Sov. Phys. JETP 46 (1977) 641;
G. Altarelli and G. Parisi, Nucl. Phys. B126 (1977) 298. 

\bibitem{Jaffe} R. L. Jaffe, Phys. Rev. Lett. 50 (1983) 228;
F. E. Close, R. G. Roberts, and G. G. Ross, Phys. Lett.
B129 (1983) 346.

\bibitem{Zhu} W. Zhu and J. G. Shen, Phys. Lett. B219 (1989) 107; 
W. Zhu and L. Qian, Phys. Rev. C45 (1992) 1397.

\bibitem{IZ2} D.~Indumathi, W. Zhu, Dortmund University preprint
DO-TH-96/12, to appear in the Proceedings of the Workshop, ``Future
Physics at HERA'', DESY, Hamburg, 1996.

\bibitem{Mucklich} A. M\"ucklich, Ph. D. Thesis, Ruprecht-Karls
Universit\"at, Heidelberg, 1995. 

\bibitem{pol} W. Melnitchouk, G. Piller, and A.W. Thomas, Phys. Lett.
B346 (1995) 165;
L.D. Kaptari {\it et al.}, Phys. Lett. B321 (1994) 271;
H. Khan and P. Hoodbhoy, Phys. Lett. B298 (1993) 181;
M.V. Tokarev, Phys. Lett. B318 (1993) 559;
B. Badelek and J. Kwiecinski, Nucl. Phys. B370 (1992) 278;
L.L. Frankfurt and M.I. Strikman, Nucl. Phys. B405 (1983) 557;
L.L. Frankfurt, V. Guzey, and M.I. Strikman, Tel Aviv
University, Israel, preprint 1996, hep-ph9602301; 
R.M. Woloshyn, Nucl. Phys. A496 (1989) 749;
C. Ciofi degli Atti, E. Pace, G. Salme, Phys. Rev. C46 (1992) R1591; 
C. Ciofi degli Atti, S. Scopetta, E. Pace, G. Salme,
Phys. Rev. C48 (1993) R968. 

\bibitem{E142} 
P.L. Anthony, et al., SLAC-E142 Collaboration, Phys. Rev. Lett. 71 (1993)
959; 
data from C.C. Young, ``Measurements of the Neutron Polarised structure
function at SLAC'', Talk presented at the International workshop on DIS
and related subjects, Eilat, Israel, Feb 6-11, 1994; See also
J. Ashman {\it et al.}, EMC, Nucl. Phys. B238 (1989) 1; 
D. Adams {\it et al.}, SMC, Phys. Lett. B329 (1994) 399; Erratum  B339
(1994) 332; 
K. Abe {\it et al.}, SLAC-E143 Collaboration, Phys. Rev. Lett. 74 (1995)
346, and preprints SLAC-PUB-94-6508 and SLAC-PUB-95-6734. 

\bibitem{I} D. Indumathi, Phys. Lett. B374 (1996) 193.

\bibitem{GRVpol} M. Gl\"uck, E. Reya, and W. Vogelsang, Phys. Lett.
B359 (1995) 201. 

\end{thebibliography}
\end{document}